\documentclass[prb,aps,epsf,twocolumn,notitlepage,showpacs,10pt]{revtex4-1}
\usepackage{graphicx,amsfonts,times,bm,amsmath, amssymb, verbatim,color,array,dsfont,hyperref}
\usepackage{appendix}
\newcommand{\bra}[1]{\langle {#1} |}
\newcommand{\ket}[1]{| {#1} \rangle}

\newcommand{\ua}{\uparrow}
\newcommand{\da}{\downarrow}
\newcommand{\ra}{\rightarrow}

\renewcommand{\>}{\rangle}
\renewcommand{\(}{\left(}
\renewcommand{\)}{\right)}
\renewcommand{\[}{\left[}
\renewcommand{\]}{\right]}
\newcommand*{\bigchi}{\mbox{\Large$\chi$}}% big chi
\usepackage{xcolor}
\definecolor{mauve}{rgb}{0.58,0,0.82}
\begin{document}

\title{Discrimination of correlated and entangling quantum channels with selective process tomography}
\author{Eugene Dumitrescu$^{1,2}$ and Travis S.~Humble$^{1,2}$}

\affiliation{ $^1$Quantum Computing Institute, Oak Ridge National Laboratory, Oak Ridge, TN 37831\\
$^2$ Bredesen Center for Interdisciplinary Research and Graduate Education, University of Tennessee, Knoxville, TN 37996}

\begin{abstract}
The accurate and reliable characterization of quantum dynamical processes underlies efforts to validate quantum technologies, where discrimination between competing models of observed behaviors inform efforts to fabricate and operate qubit devices. We present a novel protocol for quantum channel discrimination that leverages advances in direct characterization of quantum dynamics (DCQD) codes. We demonstrate that DCQD codes enable selective process tomography to improve discrimination between entangling and correlated quantum dynamics. Numerical simulations show selective process tomography requires only a few measurement configurations to achieve a low false alarm rate and that the DCQD encoding improves the resilience of the protocol to hidden sources of noise. Our results show that selective process tomography with DCQD codes is useful for efficiently distinguishing sources of correlated crosstalk from uncorrelated noise in current and future experimental platforms.  
\end{abstract}

\maketitle

\section{Introduction}
Recent multi-qubit experiments have reinforced the need to precisely characterize the dynamical processes governing emerging quantum computing devices \cite{IBM_15, Martinis_15, Monz2016, Debnath2016}. 
The convenient assumption that a qubit experiences only independent noise is rarely valid and capabilities to accurately differentiate between separable and correlated quantum dynamics is needed \cite{Gessner2015,Breuer2016}. 
The task of differentiating between two possible models for a quantum process can be cast as a decision problem in the context of statistical hypothesis testing \cite{Chiribella_PRL, Piani_PRL, Matthews_PRA, Cooney_2016}. 
In general, channel discrimination selects a model for an underlying dynamical process by inferring the completely-positive, trace-preserving map that takes a set of known input states to a set of measured output states. Measurements of the output state indirectly reveal characteristics of the CPTP map that can be used to discriminate between different potential models.
\par
Previously, quantum process tomography (QPT) has been used to completely characterize and, therefore, discriminate one- and two-qubit channels using full reconstruction of the governing quantum process \cite{Poyatos, Chuang_97, O_Brien, Bialczak, Kim, Altepeter}.
However, this complete form of QPT quickly becomes intractable for higher dimensional systems because the number of required measurements scales exponentially with the system size. Several alternative characterization methods have emerged to address the outstanding challenges of QPT such as state preparation and measurement (SPAM) errors. This includes randomized benchmarking, which reports an averaged fidelity for known gates \cite{Gaebler_PRL,Merkel_PRA}, and gate-set tomography, which requires even more measurements than standard QPT \cite{RBK_black_box,RBK_2016}. While these methods are operationally more robust for channel characterization, they are not intended for efficient channel discrimination.
\par
Complete reconstruction of a quantum process is not generally necessary for purposes of channel discrimination. Indeed, relatively few measurements may suffice to decide between different models for a set of observed behaviors. This is especially relevant for multi-qubit models, where exhaustive measurements are intractable but \textit{a priori} information about the expected dynamics may be available. We address channel discrimination in this context by using selective process tomography. In particular, we leverage recent advances in the direct characterization of the quantum dynamics (DCQD) to impose the structure of quantum error detection codes on the task of discriminating between different channel models. As shown previously, DCQD allows for selective retrieval of tomographic information characterizing a quantum process \cite{Mohseni_PRL, Mohseni_PRA, Mohseni_2010, Nigg_PRL,Graham_PRL}. 
Initially, DCQD was shown to enable piece-wise reconstruction of a channel by directly measuring elements of the underlying process matrix. 
This idea method was later extended to simultaneously encode logical qubits while performing tomographic measurements \cite{Omkar1, Omkar2}. More recently, the inclusion of error detection techniques was shown to further improve estimation of the process matrix elements in the presence of quantum noise \cite{Dumitrescu_PRA}. A similar idea has been put forward by Unden et al.~for metrological measurements \cite{Unden_2016}.
\par
We show that selective characterization of an unknown process matrix using DCQD is sufficient to perform channel discrimination. In addition, we show that DCQD codes afford a natural and transparent framework for this task while also increasing the resiliency to unknown (i.e., not modeled) channel noise. We illustrate these points using numerical simulations of multi-qubit dynamical processes under the influence of correlated and uncorrelated noise models. In particular, we consider the case of discriminating between a coherent entangling channel and its noisy equivalent. We also treat the case of selecting between incoherent correlated dynamics and an identical, independent noise model. For both examples, we confirm that selective process tomography is sufficient to discriminate the correct channel with very high probability at low false alarm rate. The latter quantities relate the performance of the protocol to operational goals.
\par
The remainder of this work is organized as follows:
in Sec.~\ref{sec:DCQD} we introduce the direct characterization methods employed to perform selective  process tomography. 
Next we formulate statistical estimation and inference tests taking the select tomographic data as input. 
Sec.~\ref{sec:cxtheta} introduces a specific example of a CNOT gate parameterized by an angle $\theta$ quantifying the amount of entanglement which can be generated by the gate. 
We numerically simulate the estimation protocol and detail the statistical process underpinning the hypothesis testing protocol. 
We further explore the efficiency of the protocol in the presence of noisy quantum sources and investigate noise filtering using quantum error detection protocols.
Next, in Sec.~\ref{sec:correlated} we explicitly address another example, namely that of estimating the degree of correlation present in a two-qubit incoherent noise source. 
We develop a model for the correlated noise and show that a two-qubit correlated channel can be detected, even in the presence of single qubit noise sources. 
Finally, we discuss the results in a broader context and make a few concluding remarks.
\section{Direct characterization of quantum dynamics}
\label{sec:DCQD}
\begin{figure}[tb]
\begin{center}
\includegraphics[width=\columnwidth]{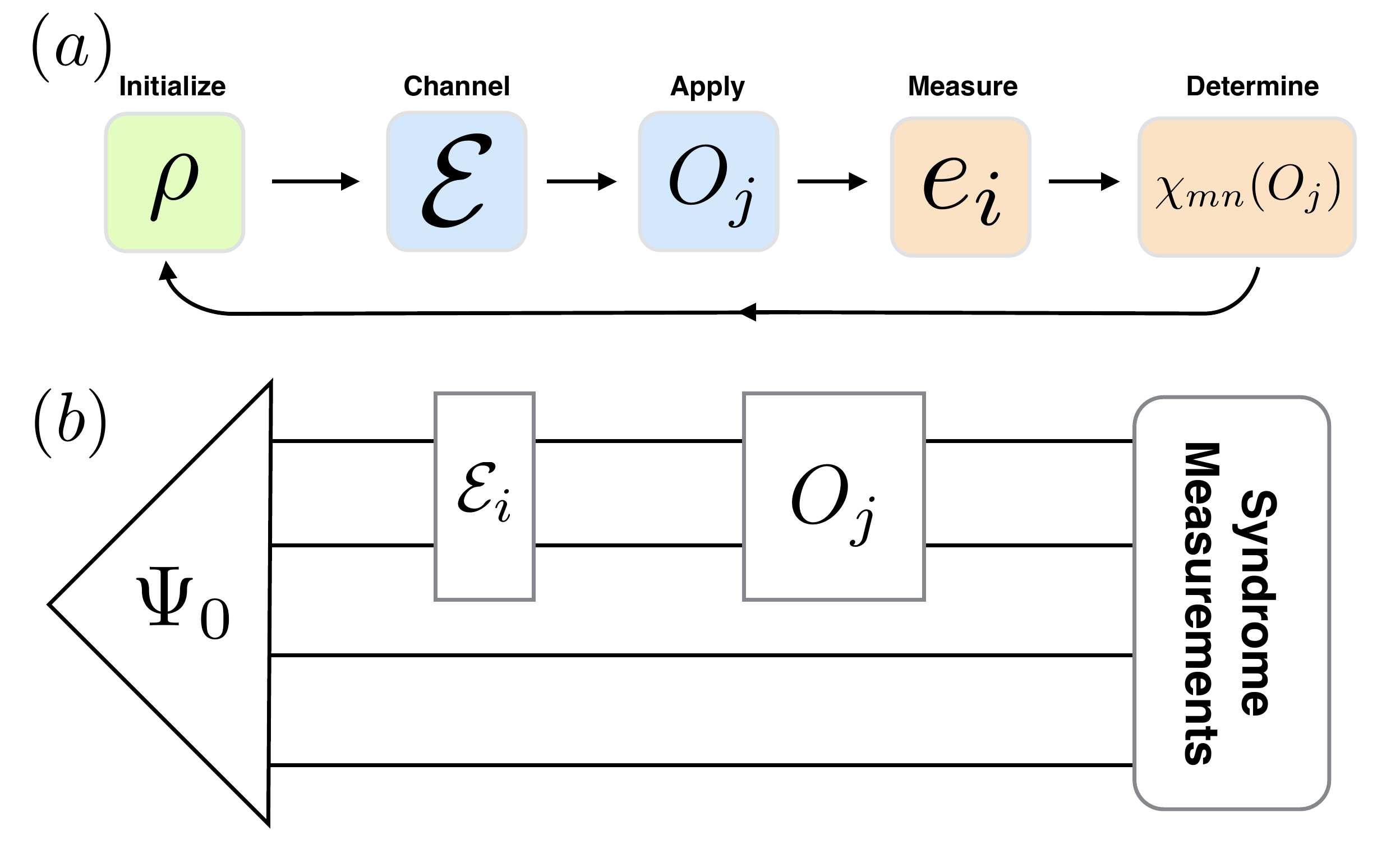}
\caption{Schematic representation of the DCQD protocol. 
The initial composite state $\rho$ (alternatively a pure state $\psi_0$) for the pure and ancilla systems evolves under an unknown channel $\mathcal{E}$. Channel discrimination selects between possible models for the channels, e.g., $\mathcal{E}_i = (\mathcal{E}_0, \mathcal{E}_1)$. 
Afterwards a unitary or projection operator may be applied to the principal system and this is followed by a syndrome readout. 
Using the syndrome measurement results as an input, the selected elements of the process matrix $\chi_{nm}$ are reconstructed.}
\label{fig:scheme}
\end{center}
\end{figure}
\begin{table}[tb!]
\centering
\begin{ruledtabular}
\begin{tabular}{cccccc}
      $i$ & $E_i$ & $\bm{e}_i$ &$i$ & $E_i$ & $\bm{e}_i$ \\[3pt]
      \hline
$0  $ & $ \mathds{1} \mathds{1}$ & $ (0  0)  0  0  0  0  $ & $8 $ & $ Z \mathds{1} $ & $ (0  0)   1   0   0   0 $ \\
$1 $ & $ \mathds{1} X $ & $(0  0)   0   0   0   1		$ & $9 $ & $ Z X $ &  $ (0  0)   1   0   0   1 $ \\
$2 $ & $ \mathds{1} Z $ & $ (0  0)   0   0   1   0 		$ &$10 $ & $ Z Z $ & $(0  0)   1   0   1   0 $\\
$3 $ & $ \mathds{1} Y $ & $ (0  0)  0   0   1   1 		$ &$11 $ & $ Z Y $ & $ (0  0)   1   0   1   1 $\\
$4 $ & $ X \mathds{1} $ & $ (0  0)   0   1   0   0 		$ &$12 $ & $ Y \mathds{1} $ & $ (0  0)   1   1   0   0 $\\
$5 $ & $ X X $ & $(0  0)   0   1   0   1				$ &$13 $ & $ Y X $ & $ (0  0)   1   1   0   1 $\\
$6 $ & $ X Z $ & $(0  0)   0   1   1   0 				$ &1$4 $ & $ Y Z $ & $(0  0)   1   1   1   0 $\\
$7 $ & $ X Y $ & $(0  0)   0   1   1   1 			$ & $15 $ & $ Y Y$ & $ (0  0)   1   1   1   1 $\\
\end{tabular}
\end{ruledtabular}
\label{tab:located_errors}
\caption{The [[4,0,2]] code error syndromes ${\bf e}_i$ partition the Hilbert space into the direct sum of states $\ket{i} = E_i \ket{0}$ indexed by the integer $i$ for the group of 16 located errors $\mathds{E}_l = \mathcal{P}_2/ \{ \pm1,\pm i \}$. 
The one-to-one correspondence between the located error operators and the syndromes means that the code is non-degenerate with respect to the set of located errors. 
The [[4,0,2]]  qubit syndromes are generalized to [[6,0,2]] qubit syndromes by the addition of the parenthesis terms as described in Ref.~\citenum{Dumitrescu_PRA}.
Unlike the errors in this table, the syndromes for weight one operators with support on the ancilla system ($\mathds{E}_u \in \mathds{E}$) begin with either $01,10$ or $11$.}
\end{table}
\par
We outline the basic theory underpinning the tomography based channel discrimination scheme. 
The central concepts presented include A) DCQD-style selective process tomography, B) model-specific parameter estimation, and C) statistical hypothesis testing. 
\subsection{Selective Process Tomography}
\label{subsec:SPT}
The DCQD protocol enables direct experimental characterization of unknown process matrix elements \cite{Mohseni_PRL, Dumitrescu_PRA}.
Underlying this idea is the Choi-Jamio\l{}kowski isomorphism between $d$-dimensional channels and $d^2$ dimensional states, which allows one to directly associate experimental measurement probabilities with process matrix elements. 
Consequently, performing DCQD tomographic measurements on an $n$ qubit system requires a minimum of $2n$ qubits. 
The first $n$ qubits form the principal system {\bf P} to be characterized while the remaining qubits represent an ancilla system {\bf A} used for measurements. 
Focusing on a two-qubit channel, consider the composite system to consist of four qubits. An example of the corresponding DCQD protocol for the two-qubit channel is presented in Fig.~\ref{fig:scheme}. 
\par
All of the qubits are initialized into a maximally entangled state with respect to the principal-ancilla bipartition, i.e., $\ket{\Psi_0} = 1/\sqrt{n}\sum_j \ket{j}\otimes \ket{j}$, where $\ket{j}$ runs over the $n$ qubit basis set. 
We assume that the initial state preparation is {\em ideal} and the density operator representation of the initial state is $\rho_0 = \ket{\Psi_0}\bra{\Psi_0}$.
It is important to note that the state $\ket{\Psi_0}$ represents the one-dimensional codespace for a stabilizer code. 
For example, the state $\ket{\Psi_0} = 1/2(\ket{0000}+\ket{0101}+\ket{1010}+\ket{1111})$ is the code state of the [[4,0,2]] error detecting code. The [[4,0,2]] code is defined by the stabilizer group $S_{[[4,0,2]]} = \<XIXI,IXIX,ZIZI,IZIZ\>$, 
where $X,Z$ denote the single-qubit Pauli bit and phase flip operators and $\ket{\Psi_0}$ is expressed in the computational basis. 
More generally, the $2n$-qubit DQCD code is generated by a group of stabilizer elements with each element having a matching support on one qubit from both the principal and ancilla systems.
To simplify notation we perform a unitary rotation to the stabilizer basis which is indexed by an integer representation of the error syndrome (i.e. $\ket{i} = E_i \ket{0}$ for $E_i$ in the Pauli group supported by the principal system, see also Tab.~\ref{tab:located_errors}). 
In the stabilizer basis the code state is simply $\rho_{0} \equiv \ket{0}\bra{0}$.  
\par
The prepared code state is next sent through the channel $\mathcal{E}_i$ as shown in Fig.~\ref{fig:scheme}. 
Expressing the output state in the stabilizer basis, $\mathcal{E}_i(\rho_0) = \sum_{j,k}{\chi_{j,k} \ket{j}\bra{k}} \equiv \sum_{j,k}{\chi_{j,k} E_j \ket{0}\bra{0} E_k^\dagger}$, highlights the connection between the process matrix representation and the DCQD code space. 
Stabilizer measurements project the composite system into the subspaces indexed by the stabilizer error syndromes $\bm{e}=\{ e_1,e_2,...,e_{2n} \}$ where $e_i=1,0$ refers to a state belonging to the $\pm 1$ eigenspaces of the generator $g_i$.
Stabilizer measurements that correspond to the state $\ket{j}$ occur with probability $\chi_{j,j}$ and characterize the diagonal of $\chi$ in the stabilizer basis.
By comparison, the off-diagonal elements of the process matrix may be recovered by applying a (two-local) unitary or projection operator to the output state $\mathcal{E}_i(\rho_0)$ before the stabilizer measurements.
This extra rotation or projective measurement maps the off-diagonal $\chi$ elements onto the diagonal, which may then be extracted by direct measurement in the stabilizer basis. Additional details regarding these DCQD measurements have been presented previously in Refs.~\citenum{Dumitrescu_PRA,Mohseni_PRL,Omkar1}.
\subsection{Model-specific Parameter Estimation}
%characterize/model specific channel
We use the DCQD framework as a tool for discriminating between a pair of quantum channels. 
The first step in comparing two models is to expand each candidate channel in terms of its process matrix representation. 
We assume the models are parameterized by a quantity of interest which is to be estimated by model-specific process tomography. 
The representation of the process matrix for a channel can have many terms that vanish (or are independent of the channel being identified) with the parameter of interest appearing in relatively few terms. 
Knowledge assumed for the model of the channel reduces the resources needed to perform channel discrimination, in contrast to the exponential resources necessary for complete tomography of an unspecified model. Unlike arbitrary assumptions, knowledge about the expected channel behavior may be inferred from composing constituent parts or from indirect characterization of the system.
\par
In general there are many ways to probe a specific parameter. 
Furthermore, the various measurement probabilities (i.e. stabilizer measurements after possible unitary rotations or projectors are applied) will depend differently on the parameter. 
This leads us to emphasize that it is important to maximize the sensitivity of the measurements depending on the parameter being estimated. 
We minimize the estimator variance, calculated using the Cramer-Rao lower bound, by picking a measurement set which maximizes the Fisher information. 
\par
Given a measurement scheme it remains to collect data from syndrome measurements and estimate the parameter. 
There is freedom in how to estimate the model parameter(s), and we compare two different estimation techniques for the different examples presented below. 
We use the maximum likelihood (ML) estimator to find the parameter that maximizes the likelihood function of the observed data in parameter space for the case of the noisy entangling channel. 
For the incoherent noise model, we develop analytic expressions that relate syndrome frequency directly to the parameter of interest, which we denote as a direct estimator. 
While the latter analytic estimation avoids the use of maximization searching (which may fail due to local maxima), we find that finite sampling can also lead to an unphysical parameter estimate, e.g. a negative probability.  We resolve this inconsistency by simply mapping all unphysical estimates to the physical estimates in an ad-hoc manner. We find this approach works quite well in the context of correlated channel discrimination. 
\subsection{Model Selection}
Parameterization by a continuous variable generates an infinite set of channels which we partition into either {\em null} or {\em alternative} classes of channels. 
For example, in Sec.~\ref{sec:correlated} we ask if a two-local gate induces correlated bit-flip errors on the qubits supporting the gate. 
The answer must either be i) yes -- the gate induces correlated errors -- or ii) no -- the noise is of a local form.
From the point of view of channel discrimination, the exact magnitude of the errors is not the main quantity of interest.
This leads to the question of given a parameter estimate, how should one decide to accept or reject the alternative hypothesis in favor of the null case?
The answer is that the decision should be statistically motivated as explained below.
\par
Statistical inference from experimental data is common in channel discrimination and in general statistical decision theory. 
The Wald test is a parametric statistical test applicable to continuous variables (our unknown parameters will be come from a continuous space) which we will use to test the true value of a parameter based on a sample data. 
The univariate Wald statistic is given as \cite{Cowan_2011}
\begin{equation}
W = \frac{(\hat{\theta}-\theta_0)^2}{\text{Var}(\hat{\theta})},
\label{eq:Wald}
\end{equation}
where $\hat{\theta}$ is our estimator, $\theta_0$ is the parameter value under the null hypothesis, and the denominator is the variance of the estimator. 
We use the Cramer Rao lower bound for the variance in Eq.~\ref{eq:Wald} which is itself found using the Fisher information $I(\theta) =- E\[ \frac{d^2}{d\theta^2} \log \( \text{Pr}(\bm{X}|\theta) \) \]$ where the expectation value $E\[\;\]$ is taken over the stabilizer outcomes ($\bm{X}$) conditional on a given value of $\theta$.
\par
The binary decision for which quantum channel should be chosen is performed by comparing the Wald statistic to a threshold value $\lambda^*$. 
If $W > \lambda^*$ then the alternative hypothesis is selected while the null is selected only when $W \leq \lambda^*$.
We use the fact that the Wald statistic is $\chi^2$ distributed under the null model to set the value of the threshold $\lambda^*$. 
This allows us to pick a critical threshold which bounds the probability of a `false alarm' event, i.e., selection of the alternative hypothesis when the null model is true. 
The probability of detection $p_D \equiv \text{Pr}(W>\lambda^*)$ approaches unity in the asymptotic limit for any $\theta\neq\theta_0$. However, given a finite set of measurements, we may still bound the range of possible $\theta$ and perform non-trivial channel discrimination. 
\section{Coherent Noise Discrimination}
%%%%%%%
\label{sec:cxtheta}
\begin{figure}[tb!]
\begin{center}
\includegraphics[width=\columnwidth]{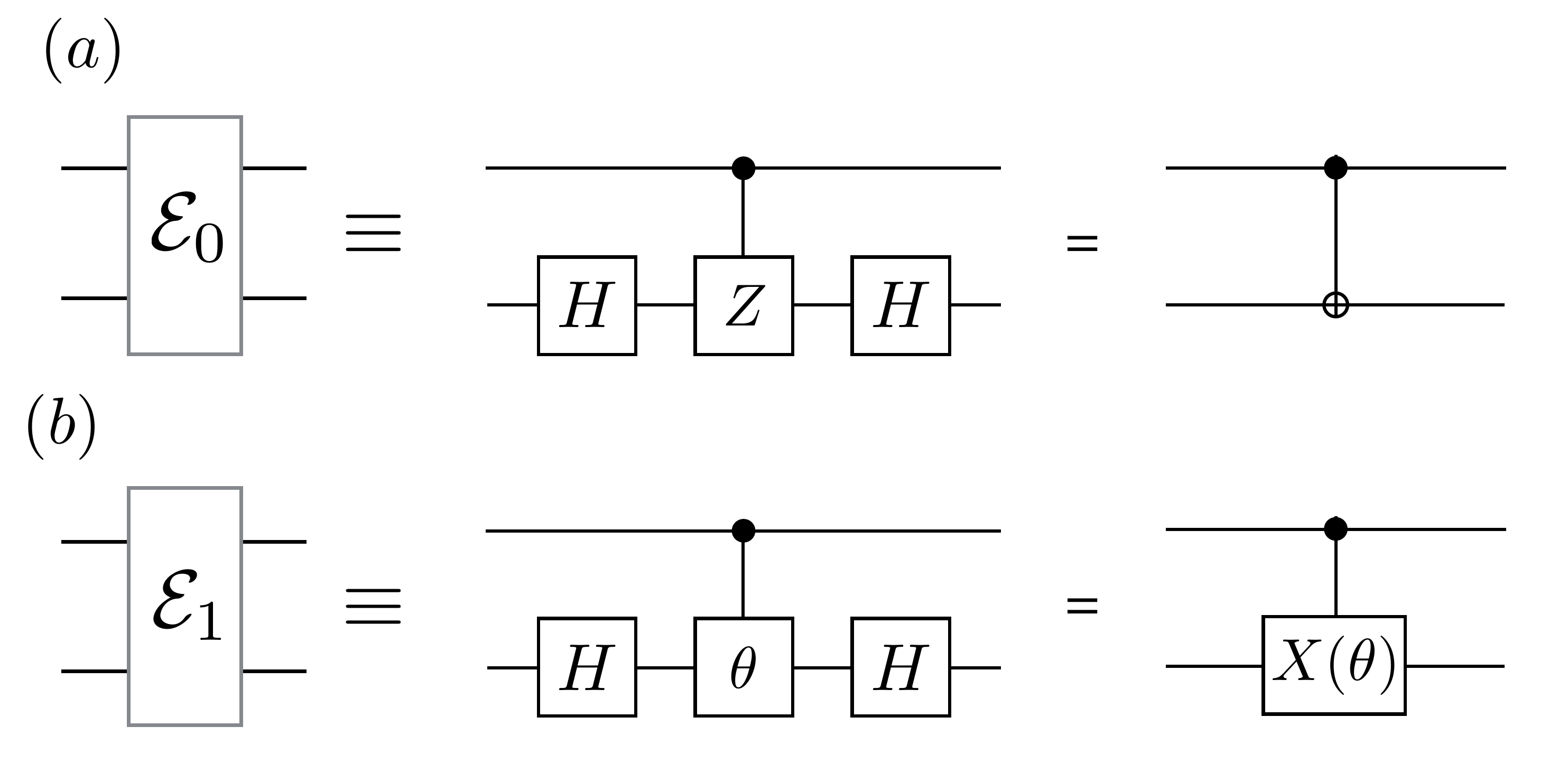}
\caption{Circuit representation of the (a) maximal and (b) noisy entangling channels denoted as $\mathcal{E}_0,\mathcal{E}_1$ respectively.
In Sec.~\ref{sec:cxtheta} we refer to $\mathcal{E}_{0(1)}$ as the null (alternative) channel. 
The process matrix decompositions of the entangling gates are given in Eqs.~\ref{eq:chi_CNOT},\ref{eq:partialCNOT}.}
\label{fig:entanglers}
\end{center}
\end{figure}
As a first example illustrating the channel discrimination, consider an imperfect CNOT that partially entangles two qubits. 
This may correspond in practice to a scenario in which an ideal CNOT gate was intended to be implemented, but evidence suggests that the resulting qubits were only partially entangled. 
We model the imperfect gate as a controlled rotation about the angle $\theta$ with the latter related directly to the degree of entanglement.
\par
We write the CNOT operator as $CX_{12} = \ket{0}\bra{0}_1 \otimes I_2 +  \ket{1}\bra{1}_1 \otimes X_2 = 1/2(I_1I_2 +Z_1I_2 + I_1X_2 - Z_1X_2)$, where qubit 1 is the control and qubit 2 is the target and the final useful expansion is a summation over operator elements belonging to the two qubit Pauli basis. 
A common way to realize the CNOT gate is via a controlled-$Z$ (phase) gate conjugated by Hadamard gates on the target qubit as shown in Fig.~\ref{fig:entanglers}(a).
Now consider a variant of the CZ gate where the $\ket{1,1}$ state acquires an arbitrary phase $e^{i \theta}$ instead of $-1$. 
This controlled phase gate interpolates between the maximally entangling CZ ($\theta=\pi$) and the trivial identity gate ($\theta =0,2 \pi$). 
Likewise, the gate generated by the controlled phase gate conjugated by Hadamard operators continuously interpolates between the identity and a CNOT as a function of $\theta$ as is represented by Fig.~\ref{fig:entanglers} (b). 
\par
The CNOT process acting on a density operator $\rho$ is $\mathcal{E}^{CX}(\rho) = CX \rho CX^\dagger$ 
%= (I_1I_2 +Z_1I_2 + I_1X_2 - Z_1X_2) \rho (I_1I_2 +Z_1I_2 + I_1X_2 - Z_1X_2)$ where we have expanded the final form in the Pauli basis.
and may also be expressed in the process matrix representation as
\begin{equation}
\label{eq:CNOT}
\mathcal{E}^{CX}(\rho)   = \sum_{m,n} \bigchi^{(CX)}_{mn} F_m \rho F_n^\dagger
\end{equation}
for which   
\begin{equation}
\label{eq:chi_CNOT}
\bigchi^{(CX)} = \frac{1}{4}\left(
\begin{array}{cccc}
 1 & 1 & 1 & -1 \\
 1 & 1 & 1 & -1 \\
 1 & 1 & 1 & -1 \\
 -1 & -1 & -1 & 1 \\
\end{array}
\right).
\end{equation}
The above matrix is represented using the partial basis $\{F^\ua_m\} = \{II,ZI,IX,ZX\}$, whose elements correspond to $E_0,E_8,E_1,E_9$, respectively, in Tab.~\ref{tab:located_errors}).
The noisy entangling gate $CX(\theta)$ is similarly expressed as $CX_{12}(\theta) = \ket{0}\bra{0}_1 \otimes I_2 +  \ket{1}\bra{1}_1 \otimes H_2 \left[\ket{0}\bra{0}_2 + e^{i \theta} \ket{1}\bra{1}_2 \right] H_2$, where $H_2  = (X_2+Z_2) / \sqrt{2}$ is the Hadamard gate acting on qubit 2.
The corresponding process matrix representation in the $\{F^\ua_m\}$ basis for the operator $CX_{12}(\theta)$ is  
\begin{widetext}
\begin{equation}
\label{eq:partialCNOT}
\bigchi^{(CX_{12}(\theta))} = \frac{1}{8}
\left(
\begin{array}{cccc}
 3 \cos (\theta )+5 & -\cos (\theta )+2 i \sin (\theta )+1 & -\cos (\theta )+2 i \sin (\theta )+1 & \cos (\theta )-2 i \sin (\theta )-1 \\
 -\cos (\theta )-2 i \sin (\theta )+1 & 1-\cos (\theta ) & 1-\cos (\theta ) & \cos (\theta )-1 \\
 -\cos (\theta )-2 i \sin (\theta )+1 & 1-\cos (\theta ) & 1-\cos (\theta ) & \cos (\theta )-1 \\
 \cos (\theta )+2 i \sin (\theta )-1 & \cos (\theta )-1 & \cos (\theta )-1 & 1-\cos (\theta ) \\
\end{array}
\right)
\end{equation}
\end{widetext}
which reduces to Eq.~\ref{eq:chi_CNOT} for the {\em null} model at $\theta = \pi$ and the identity gate for $\theta = 0,2\pi$.
With the rotated CNOT model in mind, let us now probe the entangling nature of the gate by performing a set of measurements designed to estimate $\theta$.
\par
Our goal now is to take the [[4,0,2]] code state as our resource state and perform stabilizer measurements to determine the true $\theta$ after passing through the $CX(\theta)$ channel.
The probability for each stabilizer result (denoted by $p_i$) is given by the diagonal elements $\chi_{i,i}(\theta)$ of Eq.~\ref{eq:partialCNOT}.
Relating the observed syndrome frequencies with the diagonal $\chi$ elements we can then obtain a ML estimate for the over rotation parameter $\hat{\theta}_{ML}$.
We can then use the ML estimate to represent the alternative hypothesis and use the perfect CNOT as the null hypothesis. 

The simplest measurement scheme is to make stabilizer measurements following the $CX(\theta)$ channel. 
The four non-vanishing syndrome probabilities would be $p_0=(3 \cos(\theta) + 5)/8$ and $p_1=p_8=p_9 = (1-\cos(\theta))/8$. 
Unsurprisingly, this is not the best measurement basis for estimating $\theta$. 
This is because all the stabilizer probabilities depend only on $\cos(\theta)$ which is a flat function near $\theta \approx \pi$. 
Thus, a small change in the parameter $\theta$ leads to a very small change in the measurement outcomes, in contrast to the high sensitivity one would like to achieve.
To make this observation more quantitative we calculate the Fisher information $I(\theta) =- E\[ \frac{d^2}{d\theta^2} \log \( \text{Pr}(X|\theta) \) \]$ for this set of syndrome probabilities where the expectation value $E\[\;\]$ is taken over the stabilizer outcomes conditional on a given value of $\theta$.
Note that we have verified the ML estimator is unbiased since the score of the log-likelihood vanishes for all values of $\theta$.
Our intuition is confirmed by the Cramer-Rao lower bound $\text{CRLB}(\theta) = I(\theta)^{-1}=\frac{1}{N}(\frac{2}{3 (\cos (\theta )+1)}+1)$ which fundamentally lower bounds the variance  $\text{Var}(\theta)$.
Because $CRLB(\theta)$ diverges at $\theta=\pi$ the initial diagonal scheme is not optimal and we look for another measurement basis. 

\begin{figure}[tb!]
\begin{center}
\includegraphics[width= \columnwidth]{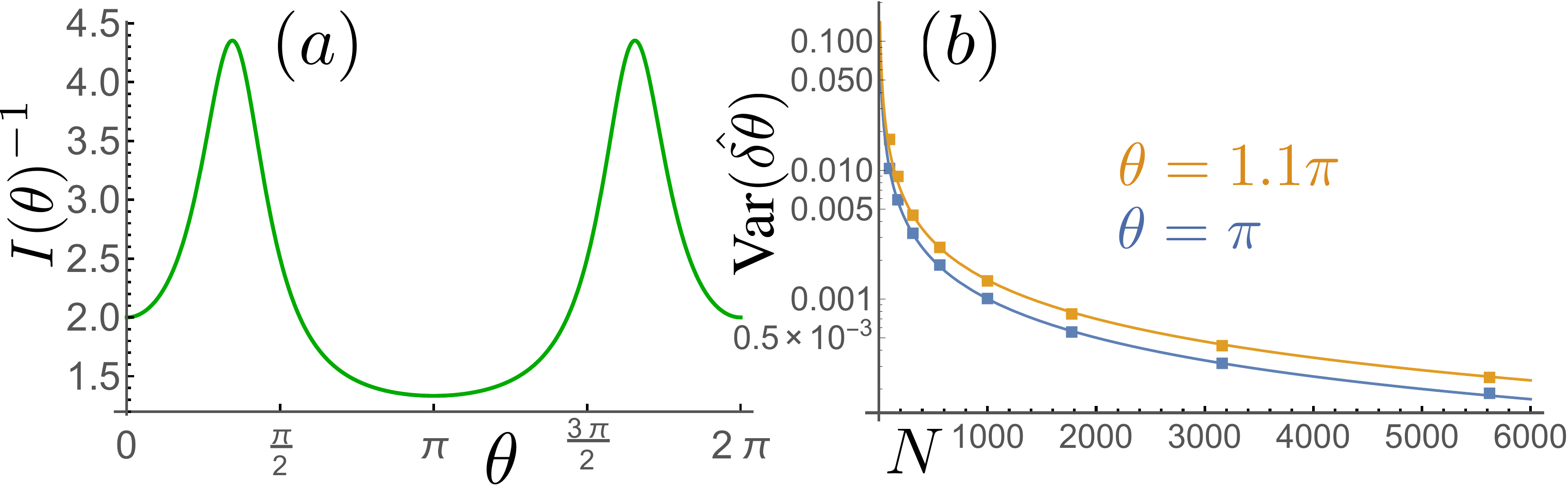}
\caption{Cramer-Rao lower bound  $\theta$ dependence for the (a) rotated measurement schemes.
(b) Numerically determined variance scaling (solid squares) of the maximum likelihood estimate $\hat{\delta\theta}$ vs. the theoretical CRLB (solid line) as a function of the number of measurements. 
Examples are provided in the case that the null ($\theta = \pi$, lower line) and alternative ($\theta = 1.1 \pi$, upper line) hypothesis underlie the data.}
\label{fig:variance}
\end{center}
\end{figure}

To increase sensitivity to variations in $\theta$, we modify the protocol by performing a pair of single qubit unitary gates, $O = U_1\otimes U_2$ in Fig.~\ref{fig:scheme}, just prior to the measurement step. 
The joint rotation $U_1 = (\mathds{1} + i Z_1)/\sqrt{2},U_2 = (\mathds{1} + i X_2)/\sqrt{2}$ is chosen because it preserves coherences between $\mathds{1}\mathds{1} \text{ and } Z\mathds{1}$ as well as between $\mathds{1}\mathds{1}\text{ and }X\mathds{1}$, both of which appear in Eq.~\ref{eq:partialCNOT} and contain $\sin(\theta)$ terms which will be sensitive to small changes in $\theta$ near $\pi$.
After the joint rotation the stabilizer outcome become, 
\begin{eqnarray}
\label{eqn:rotated_probs}
p_0 & = & (2 \sin (\theta )-\cos (\theta )+3)/8 \\ \nonumber
 p_9 & = & (-2 \sin (\theta )-\cos (\theta )+3)/8\\ \nonumber
p_1 & =& p_8= (\cos (\theta )+1)/8.
\end{eqnarray}
In terms of estimating $\theta$ the unitary rotation $U_1\otimes U_2$ turns out to be an excellent choice because the Cramer-Rao lower bound of the variance, $CRLB(\theta) = \frac{1}{N} (1-\frac{2 (\cos (\theta )-3)}{-10 \cos (\theta )+5 \cos (2 \theta )+9})$ plotted in Fig.~\ref{fig:variance} panel (a), is minimized at $\theta = \pi$. 

%%%% Paragraph related to panels (c,d) in Fig 3.
We verify that the stabilizer measurements saturates the Cramer-Rao lower bound by numerically sampling $N$ stabilizer results generating the data set $X = \{x_0,x_1,...x_{15} \}$ where each $x_i$ indicates the number of times the $i$th stabilizer outcome is measured (see Tab.~\ref{tab:located_errors}). 
Given $X$ we determine the $\hat{\theta}_{ML}$ which maximizes the likelihood function $\text{Pr}(X|\theta)$.
We repeat this calculation $M$ times to numerically calculate the variance of the distribution from which $\hat{\theta}_{ML}$ is drawn.  
In Fig.~\ref{fig:variance} panel (b) we plot the numerical variances as a function of $N$ for ture underlying values of $\theta = (\pi,1.1\pi)$ and compare to the theoretical solid lines. 
We that the lower bound is achieved for the $O = U_1\otimes U_2$ pre-processed measurements and we use the current measurement scheme for the remainder of this section.

\begin{figure}[tb]
\begin{center}
\includegraphics[width=\columnwidth]{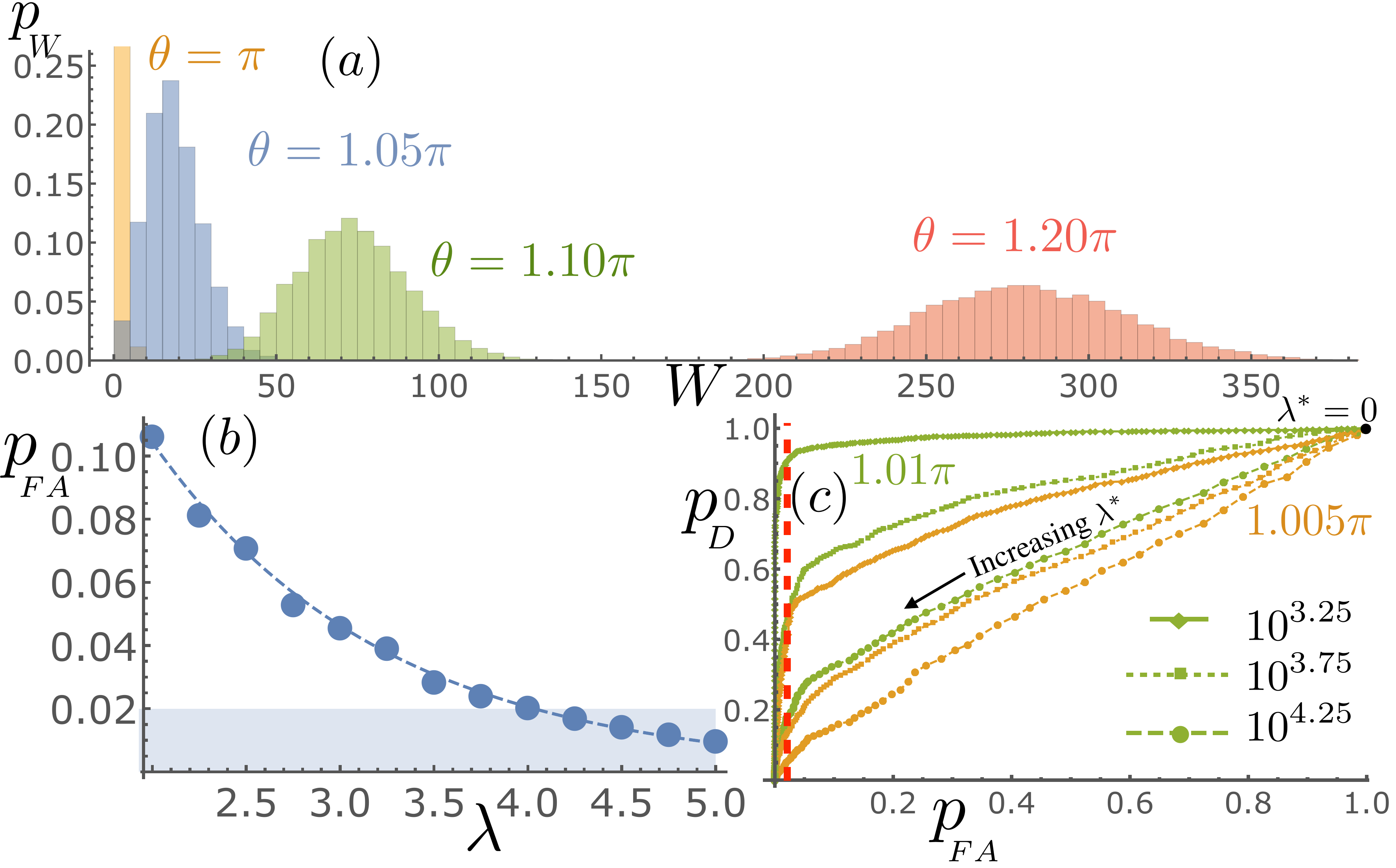}
\caption{(Color online) (a) Wald statistic distributions for various values of the true underlying $\theta$. 
(b) Numerically determined decaying probability for a false alarm detection as a function of critical test statistic $\lambda$. 
We choose $\lambda = 4$ such that $p_{FA} \leq 0.02$ as indicated by blue region.
$N=1000$ stabilizer measurements are used for each ML estimate and $2 \times 10^6$ samples are used to determine $p_{FA}$.
(c) Receiver operating characteristic curves for various underlying models ($\theta = (1.005,1.01)\pi $) as a function of the number of measurements. 
Given the $\lambda^*$ bound for $p_{FA}$, $p_{D}$ approaches unity for small $\delta \theta$ when the sample size $N$ is large.}
\label{fig:pFA_pD}
\end{center}
\end{figure}

Given a set of stabilizer data $X$ and the estimator $\hat{\theta}_{ML}$ what can be said about the {\em true} value of the parameter $\theta$? 
We answer this question by returning to the the Wald statistical test defined in Eq.~\ref{eq:Wald}.
We know that under the null hypothesis the test statistic is central $\chi^2$ distributed. 
We can use this fact to bound the probability $p_{FA}$ of a false alarm detection (i.e. the probability to reject $\theta = \pi$ when in fact the null hypothesis is true).
In Fig.~\ref{fig:pFA_pD}(b), we numerically calculate the probability $p_{FA}$ as a function of the decision value $\lambda^*$. 
In order to determine the probability that $W > \lambda^*$ we numerically create the $\chi ^2$ distribution which contains $2 \times 10^6$ samples. 
Choosing to constrain $p_{FA} <0.02$ (shaded region) we pick a critical test statistical value of $\lambda^* = 4$.
To understand the range of detectable $\theta$'s given the critical threshold chosen we turn to the receiver operating characteristic (ROC) curves plotted in Fig.~\ref{fig:pFA_pD}(c)\cite{vanTrees}.
The ROC curve shows the dependence of the false alarm vs. detection probabilities as $\lambda^*$ is varied. 
Setting $\lambda^* = 4$ simultaneously sets the false alarm probability at $p_{FA}=0.02$ (see red dashed line panel c) and restricts the range of $\theta$ one can detect given a finite number of measurements. 
Specifically panel (c) illustrates how given our choice of $\lambda^*$ modifies the probability of detecting a $\delta \theta = (0.005,0.01)\pi$ (where $\delta \theta \equiv \theta - \pi$) as a function of the number of measurements performed. 
Given $10^{4.25}$ measurements we'll detect a 1\% over rotation with unity probability and a 0.5\% over rotation with probability $p_D \approx 0.6$. In the asymptotic limit $N \rightarrow \infty$ any $\theta \neq \pi$ is detected.


\subsection{Effects of Noise}

\begin{figure*}[t]
\begin{center}
\includegraphics[width= 2 \columnwidth]{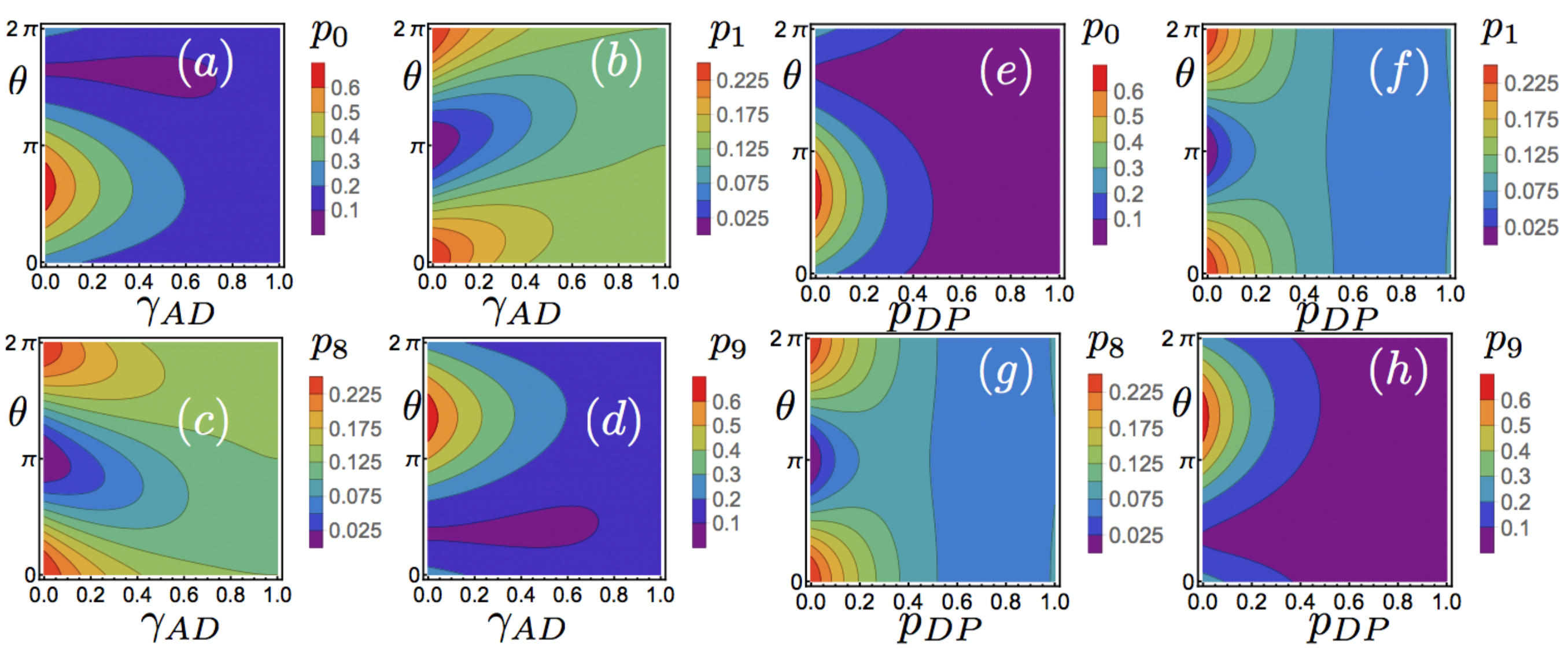}
\caption{Syndrome probabilities ($p_0,p_1,p_8,p_9$) as a function of the entangling rotation angle $\theta$ and noisy channel parameter magnitude. 
Panels (a-d) correspond to the independent and identical amplitude damping on each qubit while panels (e-h) correspond to the the depolarizing channel. 
Vertical slices at $\gamma_{AD} (p_{DP}) = 0$ correspond to the probabilities given in Eq.~\ref{eqn:rotated_probs} for noiseless measurements.
Starting from the point $\theta = \pi,\gamma_{AD} (p_{DP}) = 0$ notice the qualitatively similar behavior of the probabilities under an increase or decrease in $\theta$ while keeping $\gamma_{AD} (p_{DP}) = 0$ versus an increase in $\gamma_{AD} (p_{DP})$ while keeping $\theta = \pi$. 
The probability shift seen above coupled with the noiseless limit probability distributions Eq.~\ref{eqn:rotated_probs} bias the estimator $\hat{\theta}$.}
\label{fig:ADContours}
\end{center}
\end{figure*}

We presented in the last section how statistical inference testing can determine the character of a channel. However, employing these methods in realistic systems means accounting for the presence of noisy quantum channels and imperfect measurements in the inference test performance.   
We now investigate two different noisy quantum channels and analyze how the corresponding inference results change when using our protocol.
We expect that incorrect inferences arise because the noisy quantum channels modify the stabilizer probabilities in an unforeseen manner. 
Therefore, in an effort to mitigate the harmful effects of the noisy channels, we also consider the case where an extra pair of ancillas is used for error detection purposes as introduced in Ref.~\citenum{Dumitrescu_PRA}.
 
Consider the scheme outlined in Fig.~\ref{fig:scheme}, where a noisy quantum channel acts identically on each physical qubit. 
We assume the noisy channel acts only once and allow the magnitude of noise to vary.
Specifically, we act independently and identically with amplitude damping (AD) or depolarizing (DP) channels on each qubit. 
An operator sum representation for the amplitude damping channel on qubit $i$ is $\mathcal{E}^{AD}_i (\rho) =  \sum_{a} E_{a,i} \rho E_{a,i}^{\dagger}$ using the Kraus operators $E_{0,i} = (1 + \sqrt{1-\gamma_{AD}})\mathds{1}/2 +  (1 - \sqrt{1-\gamma_{AD}}) Z_i/2, E_{1,i} = \sqrt{\gamma_{AD}} (X_i+i Y_i)/2$. 
Likewise the depolarizing channel acting on qubit $i$ is $\mathcal{E}_i^{DP} (\rho) = (1-p_{DP}) \rho + p_{DP} (X_i \rho X_i + Y_i \rho Y_i + Z_i \rho Z_i)/3$. 

The noisy syndrome probabilities can still be calculated according to $p_i = \text{Tr}\[\Pi_i \rho \]$ but these expressions are rather complicated, so instead, to develop a qualitative understanding of how noise affects our inference decisions, we plot the probability for each of the four syndromes used for $\theta$ estimation ($e_0,e_1,e_8,e_9$) as a function of the CNOT rotation angle $\theta$ and the strength of the noise in Fig.~\ref{fig:ADContours}. 
To see how the inference is affected we focus on the syndrome probabilities in the relevant regime of small deviations $\delta\theta$ away from $\theta = \pi$ and weak noise $\gamma_{AD},p_{DP} \ll 1$. 
By increasing $\theta$ beyond $\pi$, the probabilities $p_{1},p_{8}$ increase from zero to a finite value while $p_{0}$ increases and $p_{9}$ decreases. 
The behavior of the syndromes is symmetric under the exchange of syndromes and reflection of $\theta$ about $\pi$ with $p_0(-\theta) = p_9(\theta)$ and similarly for syndromes $1$ and $8$.
The behavior under the increase in $\theta$ is qualitatively similar to that of the probabilities starting at $\theta = \pi$ and turning on the noisy channels. 
If we do not account for the noise sources in our model of the channel, the noise biases the estimator $\hat{\theta}$. 

\begin{figure}[tb!]
\begin{center}
\includegraphics[width= \columnwidth]{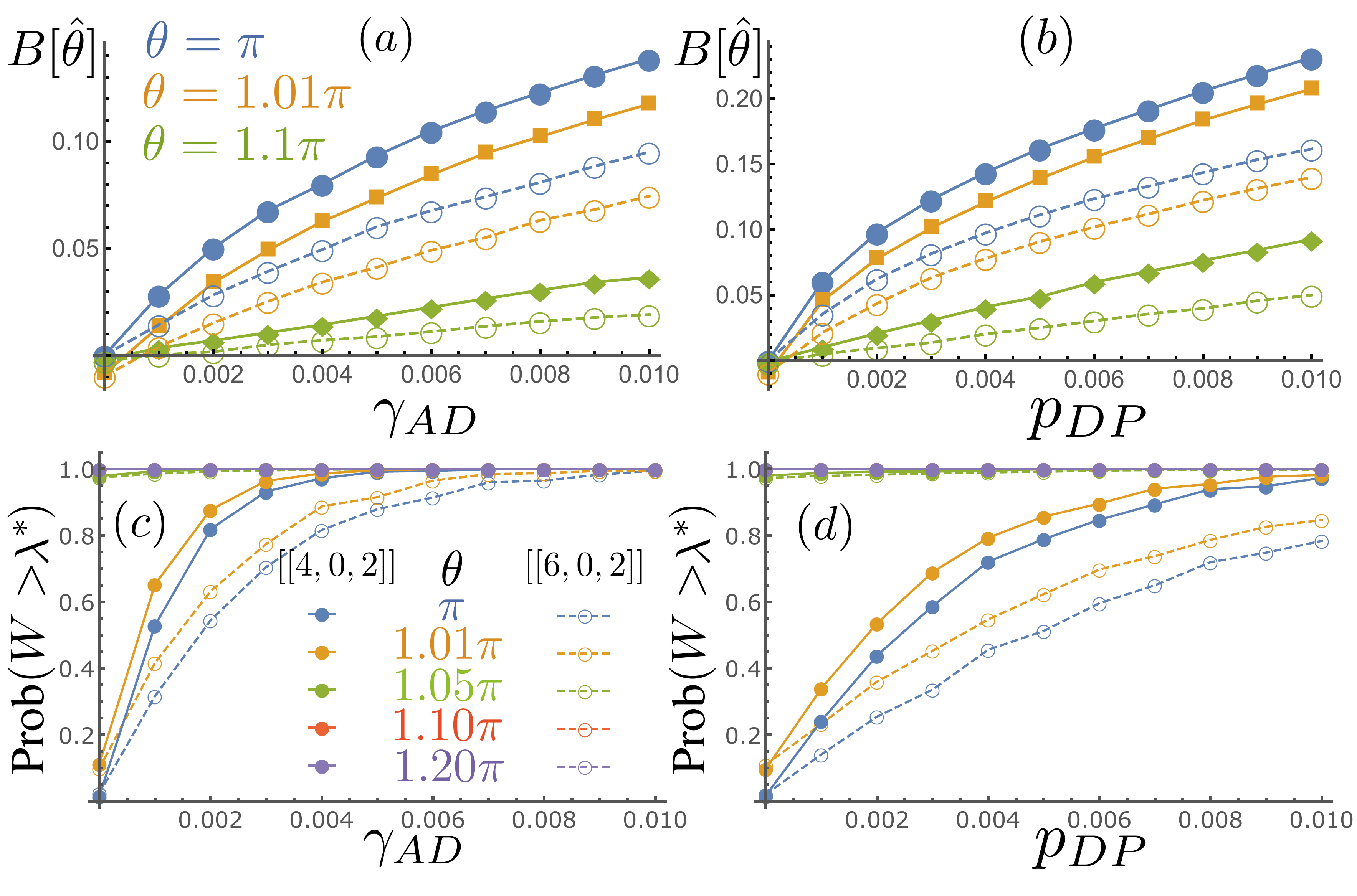}
\caption{(Color online) The estimate bias $B[\hat{\theta}]$ as a function of (a) amplitude damping and (b) depolarizing noise on the physical qubits. 
Solid lines and markers correspond to the [[4,0,2]] while dashed lines represent the [[6,0,2]].
The probability of choosing the alternative hypothesis (via the $W$ statistic) as a function of (c) amplitude damping and (d) depolarizing noise. 
The probability that $W>\lambda_{\star}$ is the probability of a false alarm, in the case that $\theta = \pi$ (blue data), and probability for a successful detection for $\theta \neq \pi$. 
Increased data selectivity due to the encoding improves the overall probability of success via a false alarm rate reduction.}
\label{fig:FailProb}
\end{center}
\end{figure}

Biasing away from $\theta = \pi$ now occurs in the presence of noise since we are actually probing an effective $CX(\theta)$ with noise channel. 
The effects of this bias are illustrated by the solid lines in Fig.~\ref{fig:FailProb} (a,b) where the solid data points correspond to the estimate bias $B[\hat{\theta}] = E[\hat{\theta}] - \theta$ as a function of noise strength.
In order to reduce the effect of this biasing, we look at a similar protocol with error detection where we utilize two ancilla qubits to generate a larger resource state known as the [[6,0,2]] code \cite{Dumitrescu_PRA} capable of detecting weight one errors on the two ancilla system qubits not involved in the CNOT rotation. 
The [[6,0,2]] code has $2^6$ syndrome outcomes, compared with $2^4$ in the [[4,0,2]] code, and is structured so that errors with support on the qubits not involved in the CNOT map the system to the newly extended syndrome space (see syndromes in parenthesis in Tab.~\ref{tab:located_errors}). 
The overall effect is that the measured syndrome probabilities can more accurately reconstruct the probabilities given in Eq.~\ref{eqn:rotated_probs}.
Indeed, this error reduction can be seen by comparing the dashed and solid data points in Fig.~\ref{fig:FailProb} (a,b).

Returning to channel discrimination, let us recall that the estimate $\hat{\theta}$ is used to make a hard decision concerning if an over/under rotation or perfect rotation has occurred. 
The decision is performed by comparing the Wald statistic, given in Eq.~\ref{eq:Wald} as a function the estimated parameter, to a critical threshold.
The overall probability for an incorrect inference is $\text{Pr}(\theta = \pi) p_{FA}+\text{Pr}(\theta \neq \pi)(1-p_D)$ where $p_{FA}$ and $p_D$ are the false alarm and detection probabilities.
Fig.~\ref{fig:FailProb} shows these quantities of interest ($p_{FA},p_D$) as function of noise strength using the $\lambda^* = 4$ threshold decision from the noiseless case discussed above. 
As expected, the internal redundancy of the $[[6,0,2]]$ code reduces $p_{FA}$ significantly for both the AD and DP channels.  

\section{Incoherent Noise Discrimination}
\label{sec:correlated}

We turn attention to the discrimination between incoherent correlated and uncorrelated sources of quantum noise. 
A correlated quantum channel acting on a multi-qubit system cannot be separated into a product of local channels. 
We express the non-separability of correlated channels as $\mathcal{E}_{i,j} (\rho) \neq \mathcal{E}_i ( \mathcal{E}_j (\rho) )$  (or the reverse order) where $i,j$ are qubit indices.
We introduce a model for an imperfect CNOT operation that is applicable to crosstalk noise. We consider an ideal CNOT gate followed by a series of one- and two-qubit, bit-flip errors.  
The task is to discriminate if the observed noise in the CNOT operation is better attributed to imperfect individual qubits or to collective correlated noise between the qubits. 
In Fig.~\ref{fig:bit_flip_channels}, we illustrates how the two model channels in this section are decomposed into uncorrelated (blue) and correlated (red) pieces.
We denote the uncorrelated noise channel as the null hypothesis and the correlated version as the alternative hypothesis.

\subsection{Correlated Noise Model}
\label{sec:BF}
%%%%%
%
\begin{figure}[t!]
\begin{center}
\includegraphics[width=\columnwidth]{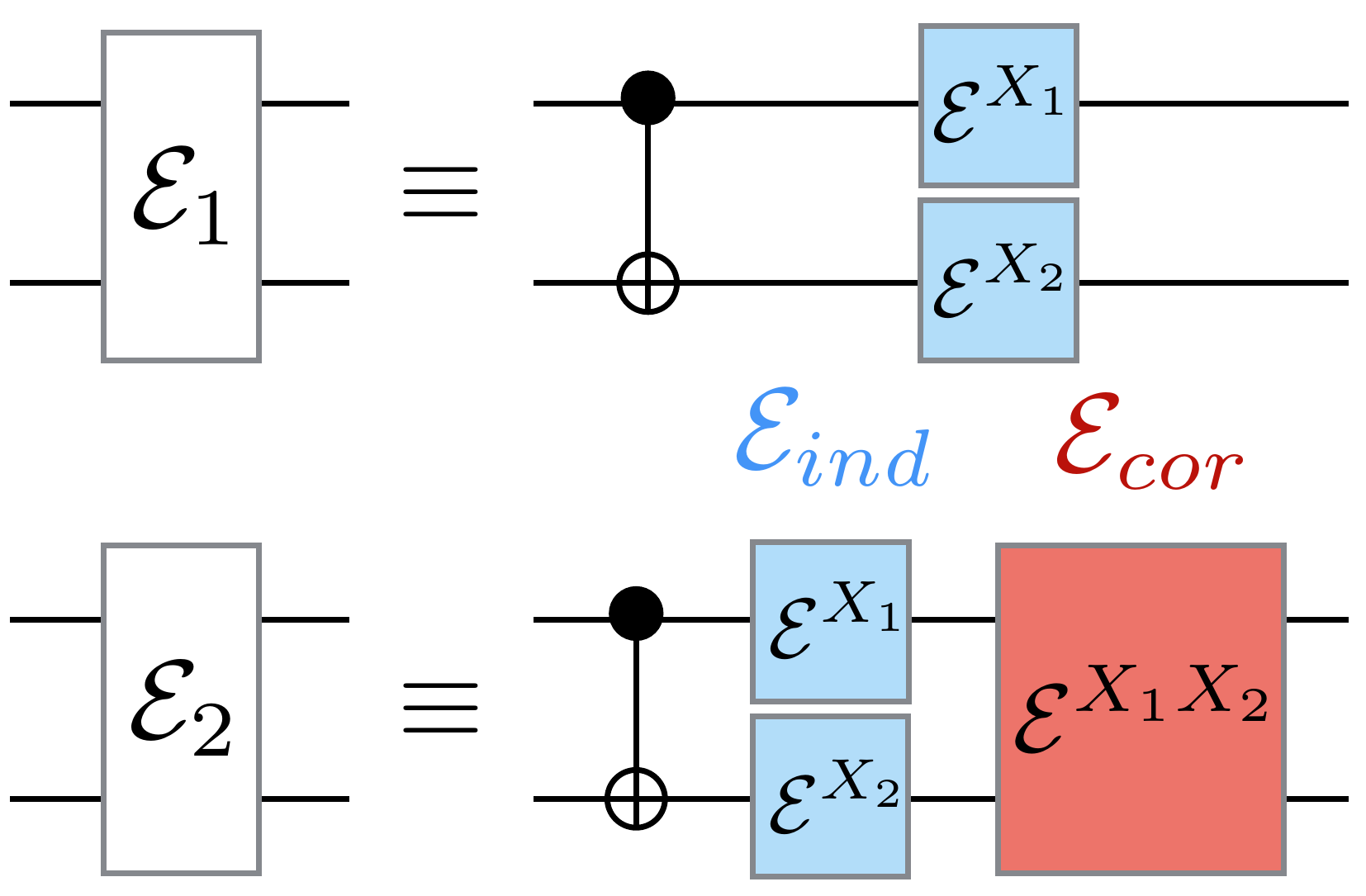}
\caption{Uncorrelated and correlated bit flip channels parameterized by the individual ($p^X_1, p^X_2$) and joint ($p^X_{12}$) bit flip probabilities.
In Sec.~\ref{sec:BF} the null (alternative) models correspond to the uncorrelated (correlated) channels.}
\label{fig:bit_flip_channels}
\end{center}
\end{figure}

The uncorrelated single qubit bit flip channel is given as $\mathcal{E}^{X}_i (\rho) = (1-p^{X}_i) \rho + p^{X}_i X_i \rho X_i$ and the correlated two qubit bit flip channel (see red box in Fig.~\ref{fig:scheme}) as $\mathcal{E}^{X}_{ij} (\rho) = (1-p^{X}_{ij}) \rho + p^{X}_{ij} X_iX_j \rho (X_i X_j)^\dagger$. 
The final state which has been subjected to a perfect CNOT followed by individual bit flips and correlated bit flips between qubits 1 and 2 can be written as (see Appendix~\ref{sec:chi_derivation} for derivation)
\begin{equation}
\label{eq:double_process}
\mathcal{E} (\rho)   = \sum_{m,n,\sigma} \bigchi^{\sigma}_{mn} F^{\sigma}_m \rho F^{\sigma \dagger}_n 
\end{equation}
where $\sigma = (\uparrow, \downarrow)$ and the summation is taken over the basis elements $F^\ua_m \in \{II,ZI,IX,ZX\},F^\da_{m} \in  \{XI,YI,XX,YX\}$.
Eq.~\ref{eq:double_process} is a general expression including the effect of both channels illustrated in Fig.~\ref{fig:bit_flip_channels}.
The form of Eq.~\ref{eq:double_process} channel is expressed in terms of a large block diagonal process matrix with blocks $\bigchi^{\ua},\bigchi^{\da}$.
In terms of the bit flip probabilities, the process matrices are
\begin{subequations}
\label{eq:chis}
\begin{eqnarray}
\bigchi^{\ua} & = & \frac{1}{4}  \left( \begin{array}{cccc} \label{eq:a}
 \alpha +\beta  & \beta -\alpha  & \alpha +\beta  & \alpha -\beta  \\
 \beta -\alpha  & \alpha +\beta  & \beta -\alpha  & -\alpha -\beta  \\
 \alpha +\beta  & \beta -\alpha  & \alpha +\beta  & \alpha -\beta  \\
 \alpha -\beta  & -\alpha -\beta  & \alpha -\beta  & \alpha +\beta  \\
\end{array} \right) \\ \label{eq:b}
\bigchi^{\da} & = & \frac{1}{4} \left( \begin{array}{cccc} 
 \gamma +\delta  & \gamma +\delta  & \gamma -\delta  & \delta -\gamma  \\
 \gamma +\delta  & \gamma +\delta  & \gamma -\delta  & \delta -\gamma  \\
 \gamma -\delta  & \gamma -\delta  & \gamma +\delta  & -(\gamma +\delta ) \\
 \delta -\gamma  & \delta -\gamma  & -(\gamma +\delta ) & \gamma +\delta  \\
\end{array} \right)
\end{eqnarray}
\end{subequations}
where we have defined the coefficients
\begin{eqnarray}
\label{eq:prob_coeffs}
\alpha & = & p_1^X p_2^X p_{12}^X+ (1-p_1^X ) (1-p_2^X ) (1-p_{12}^X ) \\ \nonumber
\beta & = & (1-p_1^X ) \left(1-p_{12}^X\right) p_2^X+p_1^X \left(1-p_2^X\right) p_{12}^X \\ \nonumber
\gamma & = & (1-p_2^X) \left(1-p_{12}^X\right) p_1^X+\left(1-p_1^X\right) p_2^X p_{12}^X \\  \nonumber
\delta & = & p_1^X \left(1-p_{12}^X\right) p_2^X+\left(1-p_1^X\right) \left(1-p_2^X\right) p_{12}^X 
\end{eqnarray}
which themselves obey the normalization condition $\alpha + \beta + \gamma + \delta = 1$.
The probability for a clean CNOT without additional $X_{1,2}$ errors is $\alpha$ while $\beta ,\gamma, \delta$ are the contributions to a CNOT with additional $X_2,X_1, X_1X_2$ operators respectively.  
Note that Eq.~\ref{eq:chis} becomes Eq.~\ref{eq:chi_CNOT} in the limit where $p_1^{X},p_2^{X},p_{12}^{X} \rightarrow 0$ (i.e. $\{\alpha,\beta,\gamma,\delta\} \ra \{1,0,0,0\}$).
Next, we'd like to relate the matrix elements of Eq.~\ref{eq:chis} to syndrome measurements, thus directly estimating each bit flip probability.

\subsection{Direct estimation of bit flip rate estimation}

We estimate the parameters for the alternative model from stabilizer measurements that relate the measured process matrix elements to the bit-flip rates.
The idea is that the $\chi$ matrix elements in Eq.~\ref{eq:chis} are to be experimentally probed via a DCQD procedure discussed in Sec.~\ref{sec:DCQD} and the matrix elements are functions of the bit flip rates via the functions $\alpha, \beta,\gamma,\delta$ defined in Eq.~\ref{eq:prob_coeffs}. 
Recalling the normalization condition $\alpha +\beta + \gamma + \delta = 1$ we have three equations for the three unknown parameters $ p^X_1,p^X_2,p^X_{12} $.
Defining 
\begin{eqnarray}
\label{eq:qrs} 
q & = & 1 - 2\alpha - 2 \beta 	\\ \nonumber
r & = & 1 - 2\beta - 2 \gamma	\\ \nonumber
s & = & 1 - 2\alpha - 2 \gamma,
\end{eqnarray}
for we find simple expressions for the bit flip parameters
\begin{eqnarray}
\label{eq:final_probs} 
p^X_1 & = & \frac{1}{2}\( 1 + \frac{\sqrt{qrs}}{s} \)  	\\ \nonumber
p^X_2 & = & \frac{1}{2}\( 1 + \frac{\sqrt{qrs}}{q} \)	\\ \nonumber
p^X_{12} & = & \frac{1}{2}\( 1 - \frac{\sqrt{qrs}}{r} \) .
\end{eqnarray}
which are implicitly functions of the $\chi$ matrix elements. 
Given stabilizer measurements, the Eq.~\ref{eq:final_probs} provide us with the alternative hypothesis estimators (i.e., with a non-zero $p^X_{12}$). 

However, in the case of the null hypothesis we set $p^X_{12} = 0$ which simplifies the null hypothesis bit flip expressions 
\begin{equation}
\label{eq:null_probs}
 p^{x}_{1} = \gamma + \delta, \;\;\;\;\; p^{x}_{2} = \beta + \delta.
\end{equation}
So unlike the maximum likelihood parameter estimation presented in the last section, we are now using the structure of the correlated bit flip model to directly estimate the hypothesis parameters through the closed form expressions given in Eqs.~\ref{eq:qrs}-\ref{eq:null_probs}. 

Now lets return to the direct estimation of the bit flip parameters. In order to do this, we'll solve Eqs.~\ref{eq:final_probs},\ref{eq:null_probs}, in terms of Eq.~\ref{eq:qrs} which will determined by finding specific elements of Eq.~\ref{eq:chis} as per the Sec.~\ref{sec:DCQD} procedure.
If we perform stabilizer measurements on the final state $\mathcal{E}^{X_1X_2}(\mathcal{E}^{X_1}(\mathcal{E}^{X_2} (\mathcal{E}^{CX}(\rho_0))))$ the probabilities for each syndrome outcome are $\bm{p} = \{ a,a,0,0,b,b,0,0,a,a,0,0,b,b,0,0\}$ where $a = \alpha + \beta  ,b = \gamma + \delta $ where $\alpha,\beta$ are defined in Eq.~\ref{eq:prob_coeffs} and $\bm{p}=(p_0,...p_{15})$.
These probabilities are simply the diagonal elements of Eq.~\ref{eq:double_process}. 
Bit flip rate estimation involves the three unknowns $p^X_1,p^X_2,p^X_{12}$ -- or alternatively any three of $\alpha,\beta,\gamma,\delta$ -- while $O_j = I$ (i.e. diagonal) stabilizer measurements only reveal the linear combinations $\alpha + \beta$ and $\gamma+\delta$. 
At least one additional measurement with an appropriate $O_j$ pre-measurement operator must is needed in order to estimate the full set of bit flip probabilities.  

Let's now find the appropriate $O_j$ operators needed to extract information about the full set of bit flip probabilities from the set of stabilizer measurements. 
Since all the off diagonal elements of Eq.~\ref{eq:chis} are real (as opposed to the imaginary terms of interest in Eq.\ref{eq:partialCNOT}) we'll need the projective operators $P^{\pm}_j = \frac{I \pm E_j}{\sqrt{2}}$ where the operators $E_j$ refer to the Pauli operators listed in Tab.~\ref{tab:located_errors} (i.e. with support on the principal system).
The set of coherence maintaining projective operators $\{ P^\pm_8 = \frac{I \pm Z_1}{\sqrt{2}} ,P_1 = \frac{I \pm X_2}{\sqrt{2}},P_9=\frac{I \pm Z_1X_2}{\sqrt{2}} \}$ will work for the task (see Ref.~\citenum{Dumitrescu_PRA} and appendix~\ref{sec:derivation} for details).  
Intuitively, the projection operators prepare a new quantum state whose syndrome probability distributions are functions of the off diagonal $\chi$ elements.
Together with the diagonal measurement data, we can completely specify $\chi^\ua,\chi^\da$. 
The projection operators $P^\pm_{1,8,9}$ can easily be implemented by measuring the single qubit eigenvalues of the $Z_1, X_2, \text{ and } Z_1X_2$ operators respectively. 
Given the state $CX_{12}|\Psi_0\rangle$ (i.e. the state before bit flips) both  $\pm 1$ outcomes (for all three single qubit Pauli projectors) occur with probability $1/2$. 
Implementing a projective operator and then measuring stabilizers, one will come up with two data sets denoted $X_\pm$, with $\pm$ denoting the measured eigenvalue of the $E_i$ used in the projector.   
The differences between the $\pm 1$ syndrome frequencies yield the process matrix elements as given by Eq.~\ref{eq:ReChis} in the appendix. 
Since only a subset of off-diagonal $\chi$ elements is needed to fully determine the bit flip probabilities we choose to use $ P^\pm_8 = \frac{I \pm Z_1}{\sqrt{2}}$ in our simulation with no loss of generality. 

As a final comment on the direct estimation method before presenting our simulation results, we have seen that one drawback of direct parameter estimation is that, given finite samples, the measured $\chi_{ij}$ can sometimes lead to unphysical parameters (with $p^X_{i}<0$). 
This issue only occurs in the regime when the true underlying model has a small $p^X_{12} \approx 0$.
In the event of an unphysical parameter we simply truncate the estimated value of $p^X_{12}$ to $0$, thus returning the correlated bit flip mapping back to a CPTP map. 
Of course, maximum likelihood and other techniques could by construction always yield physical mappings\cite{Ferrie_PRL_2016}, but the point of this section is to contrast a direct estimation technique to the ML estimation used in the previous section so we shall simply use a truncation scheme to enforce physically on the estimated dynamical mappings. 

\subsection{Results and Effects of Noise}
\begin{figure}[htbp]
\begin{center}
\includegraphics[width= \columnwidth]{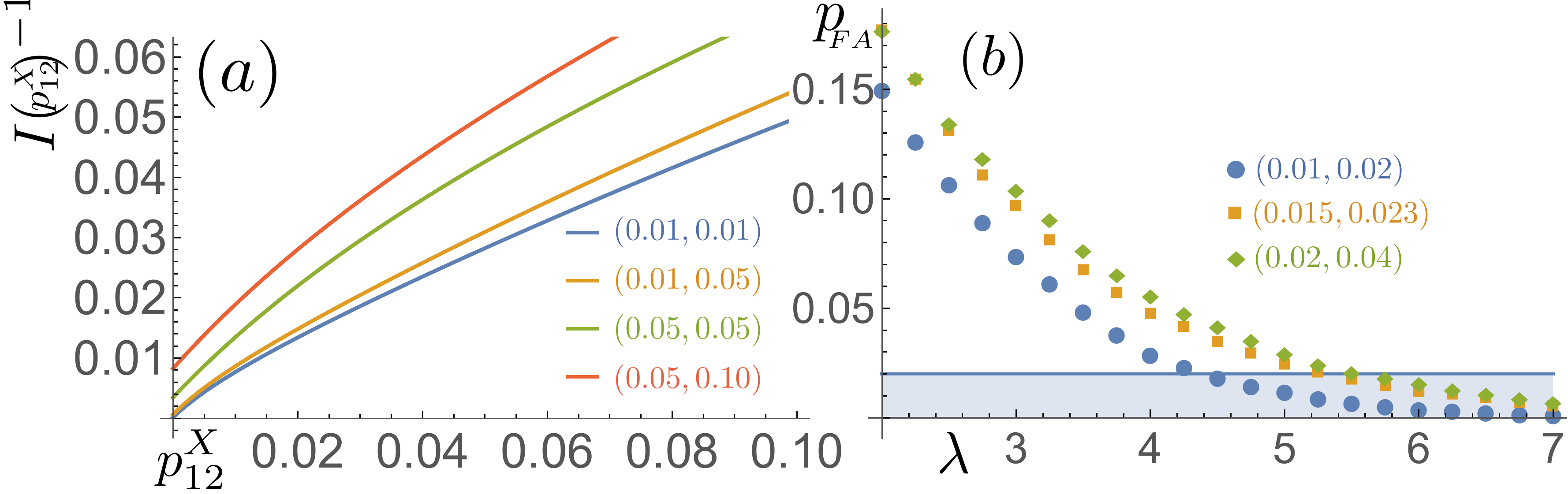}
\caption{(Color online) (a) Dependence of Cramer Rao lower-bound (normalized by number of measurements) as a function of the correlated bit flip probability. 
Each curve is indexed by the inset bit flip probability ordered pairs $(p^X_{1},p^X_{2})$. 
(b) False alarm probabilities for the Wald test as a function of the critical threshold $\lambda$. Each of the three curves also corresponds to a different set of color coded uncorrelated bit flip probabilities.}
\label{fig:FI_Thresh_BF}
\end{center}
\end{figure}
\begin{figure}[htbp]
\begin{center}
\includegraphics[width= \columnwidth]{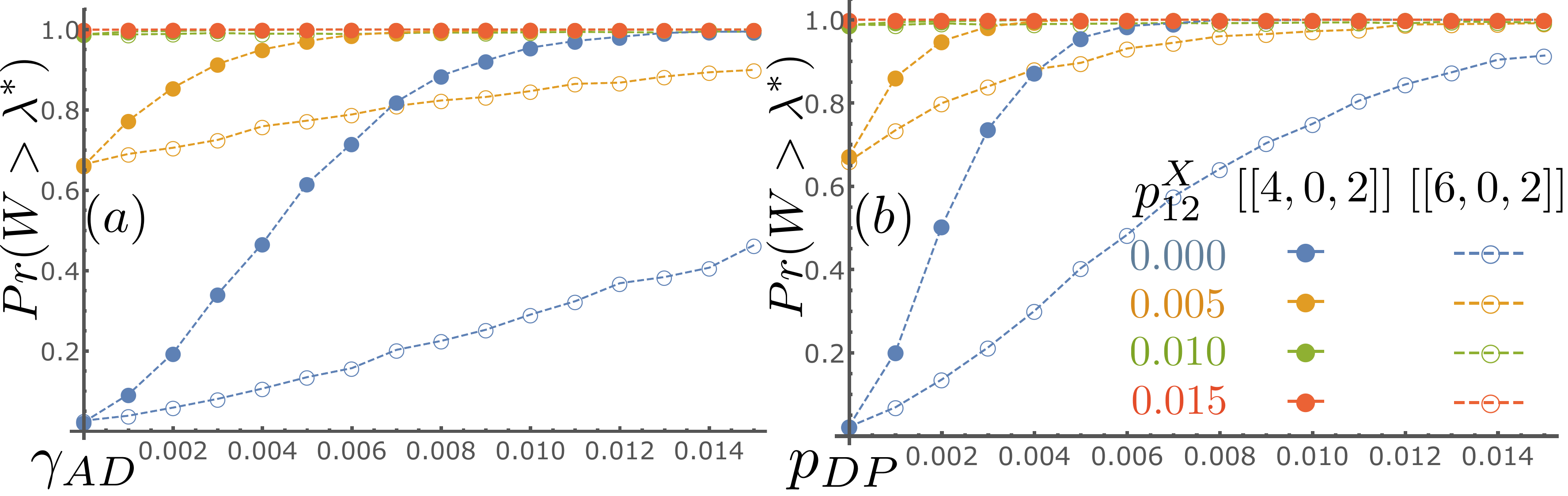}
\caption{(Color online)Panels (a,b) show the probability of the Wald statistic being greater than the critical threshold -- $p_{FA(D)}$ in the case that the null (alternative) hypothesis is true -- as a function of noisy amplitude damping (with strength $\gamma_{AD}$ ) and depolarizing channels ($p_{DP}$) strength.
True underlying values of $p^{X}_{12}$ are given by the color code in panel (b) and $p^X_1 = 0.01, p^X_2 = 0.02$ are the uncorrelated parameters.
A total of $N=1000$ measurements were used to generate a single estimation event, and a Wald statistic, and $M=5000$ instances of the simulation were used to determine the probabilities that $W>\lambda^*$.}
\label{fig:BF_fail}
\end{center}
\end{figure}

We now perform a Monte-Carlo simulation where syndrome results are sampled from a true underlying probability distribution given by  $p_i = \text{Tr}\[ \Pi_{i}  \rho \] $. 
To perform hypothesis testing, we'll again use the Wald test which takes the estimated value $\hat{p}^X_{12}$ and Fisher information as input and yields a test statistic which can be compared to a pre-determined threshold value. 
Since we are estimating all three bit flip probabilities, the Fisher information is now a matrix $I(\theta)_{ij} =- E\[ \frac{d^2}{d\theta_i d\theta_j} \log \( \text{Pr}(X|\theta) \) \]$ (which we evaluate analytically from Eq.~\ref{eq:chis}). 
The variance for each estimator is now lower bounded by the appropriate elements of the inverse Fisher information matrix. 
In Fig.~\ref{fig:FI_Thresh_BF} panel (a) we plot the Cramer Rao lower bound on $p^X_{12}$  for various $p^X_1,p^X_2$ values.
We note that the variance lower bound vanishes in the limit $\{p^X_{12},p^X_1,p^X_2\} \rightarrow 0$ and that the functional dependence on $p^X_{12}$ for the CRLB resembles the CRLB for a classical biased coin. 

Using the CRLB as a variance lower bound and an estimator $\hat{p}^X_{12}$ we can calculate the Wald statistic (Eq.~\ref{eq:Wald}) as before. 
We simulate syndrome results and calculate the distribution of $W$ given both hypothesis and various single qubit bit flip probabilities. 
The results are plotted in Fig.~\ref{fig:FI_Thresh_BF} panel (b) and a critical threshold of $\lambda*=4.25$ is chosen (for blue circled data) which sets the false alarm probability approximately to $p_{FA} = 0.02$. 
In general the threshold value depends on the independent bit flip rates as well as seen in panel (b) for three different individual bit flip rates. 
In what follows we proceed assuming the independent values chosen are $p^X_1 = 0.01, p^X_2 = 0.02$. 
 
We have so far numerically confirmed that statistical testing can be successfully implemented by directly estimating (as opposed to ML estimation) model parameters. 
Further, we have upper bounded the Wald test false alarm probability at $p_{FA} = 0.02$ by our judicious choice for the critical statistic threshold $\lambda^*$.
As before, in the event that the alternative model is true one can detect any arbitrary non-zero parameter $p^X_{12}$ in the asymptotic limit. 
To understand how our protocol performs in a realistic setting, we subject the protocol to noisy channels following the logic of Sec.~\ref{sec:cxtheta}. 
Specifically, we apply a single round of amplitude damping and depolarizing channels parameterized by $\gamma_{AD}$ and $p_{DP}$ respectively.  
We will use the probably of false alarm ($p_{FA}$)  and detection probabilities ($p_D$) as metrics for quantifying the degree to which the protocol effectiveness is reduced.
Recall that the $p_{FA}$ is the probability that $W>\lambda^*=4.25$ in the event that the null hypothesis is true ($p^X_{12} = 0$) and that $p_{FA}$ was explicitly set to $0.02$ as seen in Fig.~\ref{fig:FI_Thresh_BF}(b).  

Fig.~\ref{fig:BF_fail} (blue filled in circles) shows that $p_{FA}$ increases as the noise strength increases, eventually saturating to unity in the large noise regime.
In general, the noisy binary hypothesis probabilities $p_{FA}$ and $p_D$ for the correlated channel behave qualitatively similar to the binary hypothesis probabilities seen in Fig.~\ref{fig:FailProb}. 
However, a notable difference between the two is that for coherent rotation estimation the amplitude was more detrimental discrimination ($p_F$ increased more quickly for $\gamma_{AD}$) while for the bit flip example the situation is reversed with the depolarizing channel leading to a greater failure rate. 
The explanation for this behavior is that the depolarizing channel increases the individual bit flip rates across all qubits which directly biases the correlated bit flip parameter rate. 
On the other hand the amplitude damping channel affects other elements of the system density matrix which are not directly probed by the measurements we consider. 
Also similarly to the last section, the intuition behind the increase of $p_{FA}$ is that noisy channels modify the underlying stabilizer probabilities from the known null hypothesis probability distribution to one which resembles that of a finite $p^X_{12}$ thus biasing the estimator and increasing $p_{FA}$. 
After seeing that noise biases $p^X_{12}$ estimates upwards, it is also interesting to see how simple error detection mitigates the effects of noise. 
The empty circles in Fig.~\ref{fig:BF_fail} plot the false alarm and detection probabilities when a [[6,0,2]] \cite{Dumitrescu_PRA} initial codeword and syndromes are used.
The results of this encoding are qualitatively similar to those in the last section, however the bias reduction in this case is clearly more pronounced especially in the case of the noisy AD channels as seen in panel (a).

\section{Conclusions and Discussion}
% general conclusion
We have presented a general and intuitive theory for quantum channel discrimination based on selectice process tomography within the framework of direct characterization of quantum dynamics. This approach avoids the complete characterize of the quantum channel required by QPT for purposes of discrimination.  The main steps in the protocol include postulating (parameterized) models for the quantum channel and then selecting between these choices by comparing estimates for the model parameters. The estimates are efficient as they do not require complete tomographic reconstruction of the channel. Moreover, we have shown how unknown source of noise may bias these estimates and how the ability for DCQD codes to improve estimation lead to improved channel discrimination.
\par
% Examples in Sec III, IV
We have explicitly validated these ideas for imperfect entangling operations and for a crosstalk model involving correlated bit flips noise induced by entangling operations. Numerical simulations showed that accurate channel discrimination and parameter estimation are indeed possible, given the test statistic distribution for a detection event lies above a critical threshold value. 
This is always guaranteed in the asymptotic limit of infinite measurements, but we have also shown the capabilities for discrimination using finite numbers of measurements. 
While our simulations were made for a two-qubit principal systems, the theory itself applies to the discrimination of general multi-qubit channels. 
\par
%Noise and concatenation
The effects of unknown sources of quantum noise on discrimination have also been studied. We used amplitude damping and depolarizing channels to quantify the bias in model parameter estimates as well as the probability of an incorrect decision. In general, the addition of unknown noise is harmful to the protocol as the probability of selecting the wrong underlying model increased as a function of the noise magnitude. However, for the case of correlated bit-flip errors a single layer of DCQD error detection was shown to reduce the false alarm rate at low noise magnitudes. This result suggests that some DCQD parameter estimators are more sensitive to channel noise. 
Further, the development of an entirely self-consistent procedure to estimate specific elements of a process matrix (but not the full map) in the presence of unknown noise is deferred to future work. 
\par
% Response to referee (SPAM Discussion)
Throughout this work we have assumed the ability to reliably initialize states as well the ability to perform noiseless measurements. However, SPAM errors are present in most physical systems and therefore limit the ultimate effectiveness of tomographic protocols. In light of these issues, we conclude by discussing the impact of SPAM on the discrimination protocol as well as proposed how to potentially detect and reduce the overall effects of SPAM errors. It is also worth noting that systematic preparation and measurement errors were also partially addressed in the context of DCQD in an earlier work \cite{Mohseni_2010}.
\par
Our earlier results indirectly investigated the effect of measurement errors by adding sources of noise after the application of the channels of interest. This correspondence arises from the fact that a measurement error can be simulated by choosing the opposite syndrome measurement value with some probability. Such an error model alters the final measurement outcome probability distributions, which is exactly what our sources of error did -- albeit with a specific structure corresponding to amplitude damping or depolarizing physical processes as seen in Fig.~\ref{fig:ADContours}. State preparation measurements can analogously be included by the composition of additional noisy channels with the parameterized channel of interest. Therefore in the presence of SPAM one can expect our approach to be limited in much the same way that is summarized by the noisy results presented in Figs.~\ref{fig:FailProb} and \ref{fig:BF_fail} where the protocol is unable to distinguish between channels in the regime of large error rates, but partially retains the discriminatory capabilities for small error rates. 
\par
Using these results, one can therefore propose a simple scheme to detect the presence of SPAM. Suppose that one has control over the parameter $\theta$ in the case of the $CX(\theta)$ gate or can adjust the time between measurements to vary the depolarizing noise in the second example. One can then in principle vary the relevant parameters, traversing between the alternative and null channels, and such a change should be detected in the SPAM-less limit by the changing functional behavior of the Wald statistic on the estimator parameters as seen in Fig.~\ref{fig:pFA_pD} (a). Alternatively, in the presence of large SPAM errors one would not observe any change in the value of $W$, since $W$ would be constant as seen in Figs.~\ref{fig:FailProb} and \ref{fig:BF_fail} for the large error rate regime. 
\par
Besides the above approach for indicating the presence of SPAM errors, we would also like to develop a protocol that isolated the channel dynamics from the SPAM. While the best approach to this  task is an open question, we draw inspiration from current protocols that are designed to be robust against SPAM errors, namely randomized benchmarking and gate set tomography \cite{Gaebler_PRL,Merkel_PRA,RBK_black_box,RBK_2016}. The main idea is to repeatedly apply the channel multiple times and perform the direct characterization measurements as the number of channel applications is varied. Repeated channel application then results in an amplification of the noisy signal. For example, $N$ applications of  $CX(\theta)$ results in an over rotation angle $N \theta$ while the effect of SPAM errors is constant since there is only a single preparation and a single measurement step. Indeed this scheme actually shares some similarities to the parameter sweeping discussed above since large numbers of repeated channels would produce a single large effective parameter which could be detected despite systematic preparation and measurement errors. We are hopeful that these future modifications of the protocol will lead to a more general and robust framework of efficient channel discrimination.  

\acknowledgements
This work was supported by a grant from the Intelligence Community Postdoctoral Research Fellowship program to the University of Tennessee. This manuscript has been authored by UT-Battelle, LLC, under Contract No.~DE-AC0500OR22725 with the U.S. Department of Energy. The United States Government retains and the publisher, by accepting the article for publication, acknowledges that the United States Government retains a non-exclusive, paid-up, irrevocable, world-wide license to publish or reproduce the published form of this manuscript, or allow others to do so, for the United States Government purposes. The Department of Energy will provide public access to these results of federally sponsored research in accordance with the DOE Public Access Plan.

\onecolumngrid
\appendix
\section{Correlated Bit Flip Process Matrix}
\label{sec:chi_derivation}
Using the process matrix representation of a CNOT gate in Eq.~\ref{eq:CNOT} we see that a bit flip on the target (qubit 2) after a CNOT yields the state
\begin{equation}
\label{eq:CNOT_bf}
\mathcal{E}^X_2 (\mathcal{E}^{CX}(\rho)) = (1-p^{X}_2) \mathcal{E}^{CX}(\rho) + p^{X}_2 X_2 \mathcal{E}^{CX}(\rho) X^\dagger_2. 
\end{equation}
Noting $F_n^\dagger X_2^\dagger = (X_2 F_n)^\dagger$ we see that the second term can be re-written in the new basis $X_2 \{ F_m \} =  \{IX,ZX,II,ZI\}$ which is just a reshuffling of the basis elements with the $-1$ pre-factor factor now assigned to $ZI$. 
Using the permutation matrix for the permutation $P = \begin{pmatrix}   1 & 2 & 3 & 4 \\  3 & 4 & 1 & 2  \end{pmatrix}$ we rotate the process matrix for a $X_2$ applied to a CNOT back into the original $\{ F_m \}$ basis as
\begin{equation}
\label{eq:chi_X2}
\chi^{(CX,X_2)} = \frac{1}{4}
\left(
\begin{array}{cccc}
 1 & -1 & 1 & 1 \\
 -1 & 1 & -1 & -1 \\
 1 & -1 & 1 & 1 \\
 1 & -1 & 1 & 1 \\
\end{array}
\right).
\end{equation}
Thus, the final process matrix for a CNOT followed by a bit flip, with probability $p^{X}_2$, is a linear superposition of Eqs.~\ref{eq:chi_CNOT},\ref{eq:chi_X2} with weighting coefficients $1-p^{X}_2$ and $p^X_2$ respectively.
Repeating the above argument, but conjugating with the $X_1$ operators from the bit flips on qubit 1 channel, we can obtain a process matrix which looks like Eqs.~\ref{eq:chi_CNOT} but with matrix elements instead in the space of $\{F'_m\} = X_1 \{ F_m \} =  \{XI,YI,XX,YX\}$ operator basis elements.
Again, the process matrix with coefficients, 
\begin{equation}
\label{eq:chi_X1}
\chi^{(CX,X_1)}
 = \frac{1}{4} \left(
\begin{array}{cccc}
 1 & 1 & 1 & -1 \\
 1 & 1 & 1 & -1 \\
 1 & 1 & 1 & -1 \\
 -1 & -1 & -1 & 1 \\
\end{array} \right)
\end{equation}
Just as above, the process matrix for a CNOT and bit flip errors on qubit 1 is written in the combined $F = \{ IX,ZX,II,ZI, XI,YI,XX,YX\}$ basis as a superposition of Eq.~\ref{eq:chi_CNOT} and Eq.~\ref{eq:chi_X1} with coefficients $1-p^{X}_1$ and $p^X_1$ respectively. 
A more compact matrix representation of this channel is
\begin{equation}
\label{eq:chi_CNOT_X1_again}
\chi^{(CX,\mathcal{E}^X_1)}= \left( \begin{array}{cc}
 (1-p_1^X)\chi^{(CX)} & 0 \\
 0 & p_1^X \chi^{(CX,X_1)} \\
\end{array} \right)
\end{equation}
where each entry is a $4\times4$ matrix.

Next, we compose the {\em correlated} bit flip channel $\mathcal{E}^{X}_{ij} (\rho) = (1-p^{X}_{ij}) \rho + p^{X}_{ij} X_iX_j \rho (X_i X_j)^\dagger$ (see red box in Fig.~\ref{fig:scheme}) with the CNOT channel. 
The operator $X_1 X_2$ transforms the basis basis identically to the $X_1$ operator basis, $X_1: \{F \} \rightarrow \{F' \}$, and permutes the basis elements due to the $X_2$ operator. 
This leads to a similar expression for the process matrix of a CNOT followed by probabilistic $X_1 X_2$ errors. 
The final result is process matrix is
\begin{equation}
\label{eq:chi_CNOT_X1X2_again}
\chi^{(CX,\mathcal{E}^X_{12})}= \left( \begin{array}{cc}
 (1-p_{12}^X)\chi^{(CX)} & 0 \\
 0 & p_{12}^X \chi^{(CX,X_1X_2)} \\
\end{array} \right)
\end{equation}
where the $\chi^{(CX,X_1X_2)}$ takes the form of Eq.~\ref{eq:chi_X2} but is defined in the $\{F^\downarrow\}$ basis.
Putting together each of these individual channels we arrive at the form presented in Eq.~\ref{eq:chis}.

\section{Derivation of Eq.~\ref{eq:ReChis}}
\label{sec:derivation}

In terms of the code basis elements, the process matrices $\chi^{\ua} ,\chi^{\da}$  in Eq.~\ref{eq:chis} are indexed as
\begin{subequations}
\label{eq:chis_2}
\begin{eqnarray}
\chi^{\ua} & = & \left( \begin{array}{cccc} \label{eq:a}
 \chi_{0,0}   & \chi_{0,8}   & \chi_{0,1}   & \chi_{0,9}   \\
 \chi_{8,0}   & \chi_{8,8}   & \chi_{8,1}   & \chi_{8,9}   \\
 \chi_{1,0}   & \chi_{1,8}   & \chi_{1,1}   & \chi_{1,9}   \\
 \chi_{9,0}   & \chi_{9,8}   & \chi_{9,1}   & \chi_{9,9}   \\
\end{array} \right) \\ \label{eq:b}
\chi^{\da} & = & \left( \begin{array}{cccc} 
 \chi_{4,4}   & \chi_{4,5}   & \chi_{4,12}   & \chi_{4,13}   \\
 \chi_{5,4}   & \chi_{5,5}   & \chi_{5,12}   & \chi_{5,13}   \\
 \chi_{12,4}   & \chi_{12,5}   & \chi_{12,12}   & \chi_{12,13}   \\
 \chi_{13,4}   & \chi_{13,5}   & \chi_{13,12}   & \chi_{13,13}   \\
\end{array} \right)
\end{eqnarray}
\end{subequations}
where the index runs over the located operators listed in Table.~\ref{tab:located_errors}. 

Define $p_i (O )= \text{Tr} \[ \Pi_i O \rho O^\dagger \]$ to be the probability for the $i$th syndrome $e_i$ to be observed ($\Pi_i$ denotes the projector into the $i$th syndrome subspace) given an application of the operator $O = \frac{I \pm X_2}{\sqrt{2}}$ just prior to syndrome measurement. 
The off-diagonal coherence terms can be found as a function of the syndrome probabilities as, 
\begin{eqnarray}
\label{eq:ReChis}
Re(\chi_{0,1}) & = &  p_0(P^+_1) + p_1(P^+_1) - \[ p_0(P^-_1) + p_1(P^-_1) \]  			\\ \nonumber
Re(\chi_{4,5}) & = &  p_4(P^+_1) + p_5(P^+_1) - \[ p_4(P^-_1) + p_5(P^-_1)\]  			\\ \nonumber
Re(\chi_{8,9}) & = &  p_8(P^+_1) + p_9(P^+_1) - \[ p_8(P^-_1) + p_9(P^-_1)\] 			\\ \nonumber
Re(\chi_{12,13}) & = &  p_{12}(P^+_1) + p_{13}(P^+_1) - \[ p_{12}(P^-_1) + p_{13}(P^-_1)\]  \\ \nonumber
\\ \nonumber
Re(\chi_{0,8}) & = &  p_0(P^+_8) + p_8(P^+_8) - \[ p_0(P^-_8) + p_8(P^-_8)\]  			\\ \nonumber
Re(\chi_{1,9}) & = &  p_1(P^+_8) + p_9(P^+_8) - \[ p_1(P^-_8) + p_9(P^-_8)\]  			\\ \nonumber
Re(\chi_{4,12}) & = &  p_4(P^+_8) + p_{12}(P^+_8) - \[ p_4(P^-_8) + p_{12}(P^-_8)\]  		\\ \nonumber
Re(\chi_{5,13}) & = &  p_5(P^+_8) + p_{13}(P^+_8) - \[ p_5(P^-_8) + p_{13}(P^-_8)\]  		\\ \nonumber
\\ \nonumber
Re(\chi_{0,9}) & = &  p_0(P^+_9) + p_9(P^+_9) - \[ p_0(P^-_9) + p_9(P^-_9)\]  			\\ \nonumber
Re(\chi_{1,8}) & = &  p_1(P^+_9) + p_8(P^+_9) - \[ p_1(P^-_9) + p_8(P^-_9)\]  			\\ \nonumber
Re(\chi_{4,13}) & = &  p_4(P^+_9) + p_{13}(P^+_9) - \[ p_4(P^-_9) + p_{13}(P^-_9)\]  		\\ \nonumber
Re(\chi_{5,12}) & = &  p_5(P^+_9) + p_{12}(P^+_9) - \[ p_5(P^-_9) + p_{12}(P^-_9)\] . 		\\ \nonumber
\end{eqnarray}
Note that these definitions for the off-diagonal coherence terms are {\em independent} of the diagonal $\chi$ elements, in contrast to previous DCQD works where the off-diagonal elements require that one first solve for the diagonals \cite{Omkar1,Omkar2}.  
Experimentally determining the matrix elements relevant to our model (i.e. 16 real $\chi_{i,j}$'s) is therefore simply a matter of running the quantum circuit in Fig.~\ref{fig:scheme} (b). 

\twocolumngrid

\end{document}